\documentclass[prl,twocolumn,groupedaddress]{revtex4}
\usepackage{graphicx}
\usepackage{amsmath}
\usepackage{amssymb}
\usepackage{bm}
\usepackage{color}
\usepackage{hyperref}
\usepackage{tabularx}
\usepackage{multirow}
\usepackage{braket}
\usepackage{epstopdf}

\begin{document}
\title{Topological metals and finite-momentum superconductors}
%:Theory and application to transition metal dichalcogenides}
	
\author{{Noah F. Q. Yuan$^{1,2}$}}
\thanks{These two authors contribute equally to this work}
\email{nfqyuan@mit.edu}
\author{{Liang Fu$^{2*}$}}
%\thanks{These two authors contribute equally to this work.}
\email{liangfu@mit.edu}
\affiliation{1. Shenzhen JL Computational Science and Applied Research Institute, Shenzhen, 518109 China\\
2. Department of Physics, Massachusetts Institute of Technology, Cambridge,
Massachusetts 02139, USA}

\begin{abstract}
We show that Zeeman field can induce a topological transition in two-dimensional spin-orbit coupled metals, and concomitantly, a first-order phase transition in the superconducting state involving a discontinuous change of Cooper pair momentum. Depending on the spin-orbit coupling strength, we find different phase diagrams of 2D superconductors under in-plane magnetic field. %, which can be viewed as a projection of topological transition in three dimensions. In the corresponding superconducting phase, a first-order phase transition is realized where Cooper pair momentum changes discontinuously without symmetry breaking.
\end{abstract}
                          	
\maketitle

\textit{Introduction}---A fundamental concept in the theory of metals is Fermi surface, the locus of gapless electronic Bloch states in momentum space. While Bloch states in conventional metals are spin degenerate, the degeneracy is lifted by a magnetic field via Zeeman effect, and in non-centrosymmetric crystals, by spin orbit coupling (SOC). In both scenarios, the Fermi surface becomes spin-slit.
%In the presence of spin-orbit coupling, the Fermi surface When either symmetry is removed, spin splitting of Fermi surfaces takes place.
When attractive interaction is present at low energy, pairing instability of the Fermi surface turns a metal into a superconductor.
The interplay between spin-orbit and Zeeman splitting has interesting consequences for superconductivity, as shown in many previous works \cite{Olga,Samokhin1,Samokhin2,Patrick,Gorkov,Agterberg,
Kaur,Aoyama,Fulde,JMLu,Saito,Xi,WYHe,BTZhou,Bulaevskii,Rashba,Sigrist1,Sigrist2}.
Recent discovery of superconductivity in a variety of two-dimensional spin-orbit-coupled materials, including transition metal dichalcogenides \cite{Joe} and strontium titanate films \cite{STO}, adds new venues for further investigation of this important problem.

In this paper, we take a fresh look at spin-orbit-coupled metals and superconductors through the lens of wavefunction topology. We characterize the topology of electron wavefunctions on spin-split Fermi surfaces and establish a correspondence between topological metals in two dimensions and topological crystalline insulators in three dimensions. Applying an in-plane magnetic field to spin-orbit-coupled 2D metals can  induce a topological phase transition, characterized by a change of spin texture on the Fermi surface and $\pi$ phase shift in quantum oscillation. % and a non-analytic change of ground state spin polarization.
When the metal becomes superconducting at low temperature, the field-induced topological transition of Fermi surface is found to strongly impact electron pairing. %It can drive a first-order phase transition in the finite-momentum superconducting state, characterized by an abrupt change of Cooper pair momentum.
We present new phase diagrams of 2D superconductors under in-plane magnetic fields for various SOC strengths.
%This transition can be detected through the temperature dependence of the in-plane upper critical field. %Our theory may explain the unexpected results on superconducting strontium titanate films \cite{STO} and transition metal dichalcogenides \cite{Joe} in recent experiments.

\begin{figure}[ht]
\begin{center}
\leavevmode\includegraphics[width=1.02\hsize]{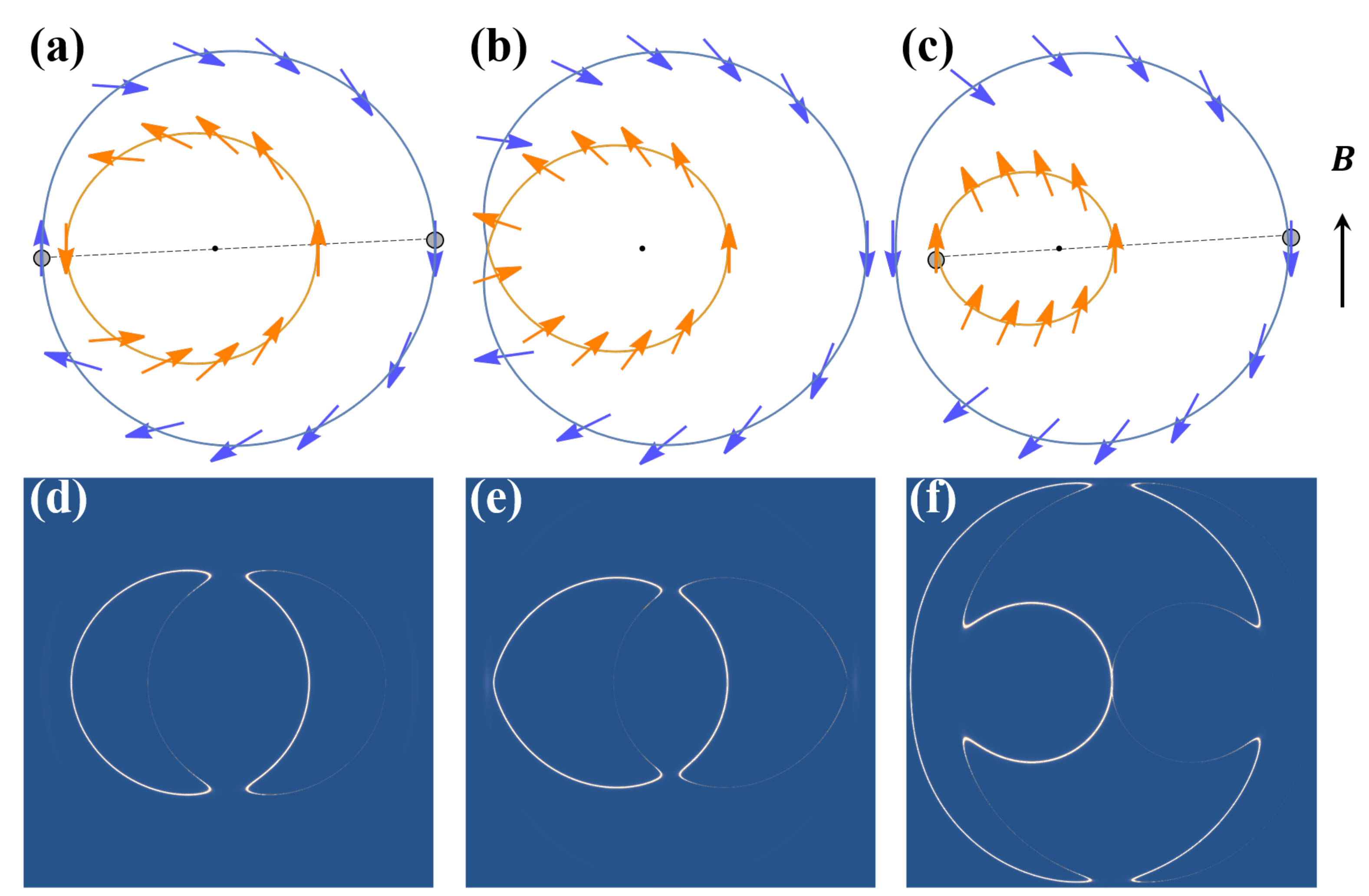}
\end{center}
\caption{(a-c) Fermi surfaces in normal phase. Yellow and blue colors denote inner and outer Fermi surfaces respectively, and arrows denote electron spins. Shaded small disks in (a) and (c) denote paired electrons in corresponding superconducting phases. (d-f) Bogouliubov Fermi segments in corresponding superconducting phase, represented by maxima of zero-energy electronic density of states (DOS) $ \rho=-\frac{1}{\pi}{\rm Im}[{\rm tr}(G\tau_{e})] $, where $G$ is Gor'kov Green's function, and $\tau_{e}={\rm diag}(1,0)$ acts in the particle-hole space.
We choose parameters $m=1,\mu=10,\alpha_{\rm R}=1,\Delta=1,$ broadening $\eta=0.01$, and magnetic field (a,d) $B=3$, (b,e) $ B=\Delta_{so}=\sqrt{20} $ and (c,f) $ B=7 $.}
\label{fig_1}
\end{figure}

This work is organized as follows. We start with a case study of 2D Rashba systems under an in-plane magnetic field, which induces a topology change of spin texture on the  Fermi surface. We then define a general set of topological invariants for 2D metals having any space-time parity symmetry in terms of quantized $\pi$ Berry phase on spin-nondegenerate Fermi surface.  %Experimental consequences of topology in metals  are predicted.
Finally we examine the impact of Fermi surface spin-splitting on superconductivity, and show that the field-induced topological transition of the Fermi surface can cause a change in pairing from intra-pocket to inter-pocket, leading to a first-order phase transition in the finite-momentum superconducting state at finite field.
{Importantly, the intra-pocket pairing state evolves smoothly from the zero-momentum Bardeen–Cooper–Schrieffer (BCS) state at zero field, while the inter-pocket pairing state evolves smoothly from the finite-momentum Fulde–Ferrell–Larkin–Ovchinnikov (FFLO) state at field beyond Pauli limit.}
By combining microscopic calculation, symmetry analysis, Ginzburg-Landau theory and physical argument, we obtain a global phase diagram of spin-orbit-coupled superconductors under Zeeman field.

%A direct cores between 2D topological metals and 3D topological crystalline insulators is made.
 %Two-dimensional topological metals can be regarded as the composition of topological surface states

\textit{Rashba systems}---
We consider a single-component 2D electron gas with SOC under magnetic field
{
\begin{equation}\label{eq_d1}
H({\bm k})=\frac{k^2}{2m}-\mu +\bm g(\bm k)\cdot\bm\sigma +\bm\sigma\cdot\bm B,
\end{equation}}
where $ \bm k=(k_x,k_y) $ is the 2D momentum, $m$ is effective mass, $\mu$ is chemical potential, $\bm g(\bm k)$ is the SOC vector, $\bm B$ is the Zeeman energy due to the in-plane magnetic field, and Pauli matrices $ \bm\sigma =(\sigma_x,\sigma_y,\sigma_z) $ denote spin.

{As a concrete example, we consider Rashba SOC $\bm g=\alpha_{\rm R}\bm k\times\hat{\bm z}$ with Rashba coefficient $\alpha_{\rm R}$.}
At ${\bm B}=\bm 0$,  in the energy eigenstate electron's spin is tied with its momentum due to Rashba SOC. As a result, two concentric Fermi circles are present, with helical spin textures of the same chirality. %To reveal the effect of Rashba SOC on superconductivity under in
%For large $\alpha_{\rm R}$, the size and DOS of the two Fermi surfaces can
The shape and spin configuration of both Fermi surfaces evolve with the in-plane magnetic field.
As  ${\bm B}$ increases, the two pockets approach each other and deform into ovals known as Cartesian ovals (Fig. \ref{fig_1}a). Electron's spin lies within the $xy$ plane and winds by $2\pi$ around both inner and outer Fermi surfaces.
At a critical field $B=\alpha_{\rm R}k_{\rm F}\equiv\Delta_{so}$ with $ k_{\rm F}=\sqrt{2m\mu} $, % is the Fermi momentum without spin-orbit or Zeeman field,
the two ovals touch each other at a point $ \bm k_{\rm P}\equiv k_{\rm F}\hat{\bm z}\times\hat{\bm B} $, where a two-fold spin degeneracy arises.
%Unlike conventional van Hove singularities in the single band, the overtitled Dirac point contributes no divergence to density of states (DOS) though there are also two crossing lines as Fermi contours near the point.
The resulting Fermi surface is a single self-intersecting curve known as a limaçon of Pascal (Fig. \ref{fig_1}b).  As the field increases further, these two ovals disconnect again and move away from each other (Fig. \ref{fig_1}c). Now, the spin winding number on each Fermi surface is $0$, and the spin configuration resembles more the Zeeman-dominated case. The inner Fermi surface shrinks further and eventually disappears at sufficiently high fields.  %$B_{\rm F}\equiv\mu+\frac{1}{2}m\alpha_{\rm R}^2$. %Usually $B\ll B_{\rm F}$.

The merging of two pockets and the change of spin winding number at $B=\Delta_{so}$ marks a new type of Fermi surface  topological transition. It is fundamentally different from Lifshitz transitions that reconnect Fermi contours through a saddle point in the energy dispersion. Here instead, the original Dirac point  at ${\bm k}=0$,  moves to $\bm k_{\rm P}$ at the Fermi level  and results in Fermi surface touching at $B=\Delta_{so}$. The dispersion around $\bm k_{\rm P}$ takes the form of an \textit{overtitled} and anisotropic Dirac cone:
\begin{eqnarray}
H({\bm k_{\rm P} + \bm p})=-v_{\rm F}p_{x}+\alpha_{\rm R}(p_{x}\sigma_y -p_y\sigma_x),
\end{eqnarray}
where $ v_{\rm F}=\sqrt{2\mu/m} $ is the Fermi velocity. Moreover, the topology of Fermi surface in momentum space is unchanged before and after the transition, and the density of states (DOS) remains finite throughout, unlike the van Hove singularity resulting from saddle points.

\textit{Topology of metals}---
Unlike Lifshitz transition associated with Fermi surface geometry, what we uncovered in Rashba systems involves the topology of quantum wavefunction on a spin nondegenerate Fermi surface. %While this topology is characterized by the spin winding number.
To characterize the wavefunction topology, we  introduce a general topological invariant for 2D systems having any parity symmetry, i.e., invariant under any transformation that reverses the orientation of space-time manifold, including time-reversal $\mathcal{T}$, reflection $M$ ($x\rightarrow -x$), and the combined operation of two-fold rotation ($x\rightarrow -x, y\rightarrow -y$) and time-reversal $C_2\mathcal{T}$.
%Since the Berry curvature is odd under parity transformation, the presence of parity symmetry enforces the $U(1)$ Berry phase on a spin nondegenerate Fermi surface that maps onto itself under parity transformation to be either $0$ or $\pi$.
{
We define the Berry phase $\varphi=\oint_{E_{\bm k}=\mu}\bm A_{\bm k}\cdot d\bm k$ as the topological invariant of a Fermi surface, where $ \bm A_{\bm k}=-i\psi^{\dagger}_{\bm k}\nabla_{\bm k}\psi_{\bm k} $ is the Berry connection and $ \psi_{\bm k} $ is the wavefunction with energy $E_{\bm k}$.
One can transform the Berry phase $\varphi$ into an integral of Berry curvature $ \Omega_{\bm k}=(\nabla_{\bm k}\times\bm A_{\bm k})_z $ on the $\bm k$-space section enclosed by the Fermi contour, namely $ \varphi=\int_{E_{\bm k}<\mu}\Omega_{\bm k}d^2\bm k $.
Under the parity symmetry, we find energy dispersion is even $ E_{\bm k}=E_{P\bm k} $ while Berry curvature is odd $ \Omega_{\bm k}=-\Omega_{P\bm k} $, where $P$ can be $\mathcal{T}, M$ or $C_2\mathcal{T}$.
As a result, $\varphi=-\varphi$({mod} $2\pi)$, and the Berry phase would be either zero or quantized to $\pi$, if inside the enclosed section there are even or odd number of Dirac points respectively.
At a Dirac point, two bands cross and the Berry curvature becomes singular.
Due to the parity symmetry, Dirac points are restricted to be at time-reversal-invariant points ($P=\mathcal{T}$) or on the reflection-invariant lines ($P=M$) or anywhere in the momentum space ($P=C_2\mathcal{T}$).
}

This quantized Berry phase therefore serves as a $Z_2$ topological invariant. Generally speaking, Zeeman splitting of Fermi surface results in  $0$ Berry phase, while spin-orbit splitting results in $\pi$ Berry phase on each Fermi surface enclosing a time-reversal-invariant momentum. In 2D Rashba model, the $Z_2$ topological distinction continues to hold in the presence of a Zeeman field, which preserves reflection and $C_2\mathcal{T}$.
{We can also include higher order SOC such as hexagonal warping \cite{warping} to break $C_2\mathcal{T}$ symmetry while preserving reflection $M$ $(x\to -x)$, so that SOC vector reads $ \bm g=\alpha_{\rm R}\bm k\times\hat{\bm z}+w(k_{+}^3+k_{-}^3)\hat{\bm z} $ where $k_{\pm}=k_x\pm ik_y$. In this case, the Dirac point is on the $k_x=0$ line as restricted by $M$, and only magnetic field $\bm B\parallel x$ can drive a topological transition of Fermi surfaces.}

We refer to Fermi surfaces having quantized  Berry phase $\pi$ and $0$ as topological and trivial  respectively. It is interesting to note that Fermi surfaces with quantized $\pi$ Berry phase are the hallmark of surface states in 3D topological (crystalline) insulators protected by parity symmetry---$\mathcal{T}$, $M$ or $C_2\mathcal{T}$ \cite{FKM, TCI,FangFu1,Sato, FangFu2}. While these topological surface states are known as ``half'' of 2D metals, we now turn this viewpoint the other way. 2D topological metals, defined as having spin-nondegenerate Fermi surfaces with $\pi$ Berry phase, can be viewed as a ``sum'' of topological surfaces. %In contrast, trivial metals only have Fermi surfaces
Thus follows a correspondence between 2D topological metals and 3D dimensional topological (crystalline) insulators.

Since the sum of Berry phases over all Fermi surfaces must be zero in any 2D metal with parity symmetry, topological metals have an even number of spin-nondegenerate Fermi surfaces with $\pi$ Berry phase. A transition from a topological metal to a trivial one generally involves the touching of two Fermi surfaces at a band degeneracy point to enable the change of Berry phase from $\pi$ to $0$ on each Fermi surface.
We thus conclude that an in-plane magnetic field generally induces a topological phase transition in spin-orbit-coupled 2D metals, provided that parity symmetry (e.g., $C_2\mathcal{T}$ or $M$) is present.

The presence of spin-orbit-split Fermi surface leads to beatings in quantum oscillation phenomena, as observed in semiconductor heterostructures and noncentrosymmetric metals. The Berry phase change from $\pi$ to $0$ across the topological transition can be further detected by analyzing the phase shift of quantum oscillation as a function of in-plane magnetic field \cite{Aris,Aris1}.  We also note recent works on topological characterization of 3D metals under Zeeman fields using Chern numbers on Fermi surfaces \cite{XiDai,Aris1}.

When attractive interaction is present, metals with spin-split Fermi surfaces may become unstable to pairing at low temperature. The competition between spin-orbit and Zeeman splitting, which drives the topological transition in the normal state, also significantly impacts the superconducting state, as we now turn to.

\textit{Finite-momentum superconductivity}---
We consider clean superconductors with a local attraction and an energy gap that is small compared to Fermi energy. In this case, superconductivity at zero field is conventional BCS type. Increasing temperature $T$ and/or in-plane magnetic field $\bm B$ drives superconducting to normal transition.
The superconducting order parameter near the transition is determined by the pair susceptibility \cite{Fulde,Sigrist1,JMLu}:
%Our When an in-plane magnetic field is applied,  change in the form of pairing is caused by the field-induced change of normal state Fermi surface. The latter depends crucially on the interplay between SOC and Zeeman field.
%Corresponding to the topological transition of Fermi surfaces in the normal state, magnetic field also drives phase transitions in the superconducting state.
%Close to second order phase transitions between superconducting and normal states, the system is governed by its pair susceptibility.
%A Cooper pair with momentum $\bm q$ is formed by two electrons with momenta $ \frac{1}{2}\bm q\pm\bm k $, which experience effective magnetic fields $\bm h_{\pm}=\bm B+\alpha_{\rm R}\hat{\bm z}\times(\frac{1}{2}\bm q\pm\bm k)$ respectively. At temperature $T$, pair susceptibility to such Cooper pairs is
\begin{eqnarray}
\chi({\bm q},\bm B,T) =N_0\log\frac{\omega_D}{T}- \sum_{s=\pm }\delta \chi_{s}({\bm q},\bm B,T),
\end{eqnarray}
where $\bm q$ is Cooper pair momentum. The first term is the $\bm q=0$ pair susceptibility in BCS theory, where $N_0$ is the total DOS of Fermi surfaces and $\omega_D$ is the Debye frequency. The second term is the correction due to Fermi surface spin splitting and finite Cooper pair momentum:
\begin{widetext}
\begin{equation}\label{eq_dchi}
\delta \chi_{s}= \oint_{{\rm FS}_{s}}\frac{dk}{|\bm v|} \left\{\phi \left(\frac{Q+s\mathcal{E}_{+}}{2\pi T}\right)\cos^2\frac{\theta}{2}+ \phi \left(\frac{Q+s\mathcal{E}_{-}}{2\pi T}\right)\sin^2\frac{\theta}{2}\right\}
\end{equation}
\end{widetext}
is contribution from inner ($s=+$) or outer ($s=-$) Fermi surface,
$\phi(x)={\rm Re}\psi\left(\frac{1+ix}{2}\right)-\psi\left(\frac{1}{2}\right),$
$\psi$ is the digamma function, $\bm v$ is electron velocity,
$Q=\bm v\cdot\bm q$ is the depairing energy of finite momentum pairing, $ \mathcal{E}_{\pm}=|\bm h_{+}|\pm |\bm h_{-}| $ is the Zeeman depairing energy of inter ($+$) or intra-pocket ($-$) Cooper pairs, $ \theta=\langle\bm h_{+},\bm h_{-}\rangle $ is the angle between $\bm h_{\pm}=\bm B+\bm g(\frac{1}{2}\bm q\pm\bm k)$, and $\bm g(\bm k)$ is the SOC vector.
%{For Rashba systems, $\bm g(\bm k)=\alpha_{\rm R}\bm k\times\hat{\bm z}$.}

In the presence of Rashba SOC, the inner $(s=+)$ and outer $(s=-)$ Fermi surface at $B=0$ have different DOS given by $ N_{s}=\frac{1}{2}N_0(1-\frac{1}{2}s\lambda) $ with $ \lambda=\Delta_{so}/\mu $, respectively. As shown in Refs. \cite{Patrick,Gorkov}, when $\Delta_{so}\gg\Delta_0$, this DOS asymmetry (often neglected) %which is sometimes neglected in previous works, %\cite{Olga,Samokhin1,Samokhin2,Fulde,Sigrist1,Sigrist2},
is important in determining Cooper pair momentum and critical fields at low temperatures. In this work, we always take into account DOS asymmetry $N_{+}\neq N_{-}$ when SOC is present.
%The free energy can be expanded as $ F=\alpha\Delta^2+\beta\Delta^4 $ in terms of the pairing potential $\Delta$, and the leading order coefficient is $\alpha=N_0\log(T/T_c)+\chi_{+}+\chi_{-}$, where $T_c$ is the zero-field critical temperature.

From pair susceptibility, the in-plane critical field $B_c(T)$ is given by $v\chi_{\rm max}=1$, where $v$ is the attractive interaction strength in $s$-wave channel and $\chi_{\rm max}$ is the maximum of $\chi(\bm q)$ among all $\bm q$. In this way we also determine the Cooper pair momentum $\bm q$ near $B_c$.

%For 2D Rashba systems $\bm g(\bm k)=\alpha_{\rm R}\bm k\times\hat{\bm z}$, depending
Depending on the magnetic field $B$ and Rashba coupling energy $\Delta_{so}$ in comparison to the BCS gap $\Delta_0=2\omega_D e^{-1/N_0v}$, we find three phases at $T=0$ near upper critical field: the normal phase (N), and finite-momentum superconducting phases whose Cooper pairs are dominantly inter-pocket (I) and intra-pocket (II) respectively.
In the limit of vanishing SOC $\Delta_{so}=0$, phase I reduces to the well-known FFLO state, which occurs at magnetic field above the Pauli limit $B_{P}\equiv\Delta_0/\sqrt{2}$ \cite{FF,LO}. Here the Zeeman splitting of the Fermi surface favors inter-pocket pairing between majority and minority spin states.
In the opposite limit of large SOC $\Delta_{so} \gg \Delta_0$, the helical spin texture of spin-orbit-split Fermi surface favors intra-pocket pairing (II). This phase can survive magnetic fields much larger than Pauli limit, and is destroyed only when Zeeman energy becomes comparable to SOC and distorts the Fermi surface significantly \cite{Patrick,Gorkov}.

\begin{figure}[ht]
\begin{center}
\includegraphics[width=1.0\hsize]{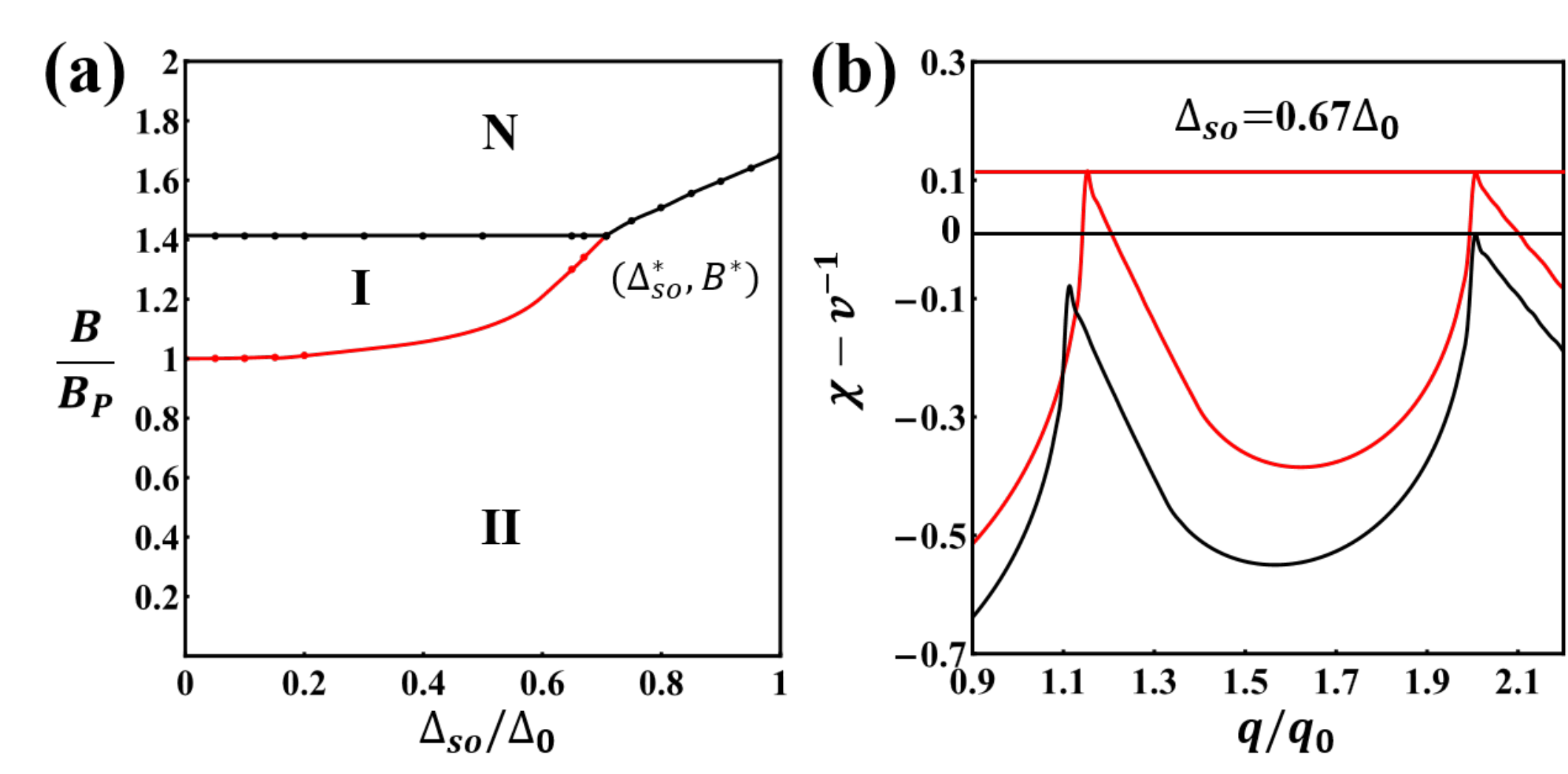}
\end{center}
\caption{(a) Phase diagram in the $\Delta_{so}$-$B$ plane at zero temperature, where N is the normal phase, I and II denote superconducting phases whose Cooper pairs are mainly inter- and intra-pocket respectively. Black (red) lines denote second (first) order phase transitions. (b) Pair susceptibility at the first (red) and second (black) order phase transitions when $\Delta_{so}=0.67\Delta_0$, where $q_0=\Delta_0/v_{\rm F}$. Dots are from numerical calculations and lines are from interpolation of dots. We set $\mu =50 T_c$ to include DOS asymmetry.}
\label{fig_2}
\end{figure}

By calculating $\chi(\bm q,\bm B, T=0)$, we obtain the phase diagram shown in Fig. \ref{fig_2}a.  Phases I and II are both finite-momentum superconductors indistinguishable by symmetry. They are separated by a first-order quantum phase transition, where the Cooper pair momentum changes abruptly mainly due to the change of pairing from inter-pocket to intra-pocket.
{As shown in Fig. \ref{fig_1}a, the topological metal $(B<\Delta_{so})$ has helical spin configuration of Fermi surfaces and hence electrons from the same Fermi surface can have opposite spins, which results in the intra-pocket pairing (II) of Fig. \ref{fig_1} d. On the contrary, the trivial metal $(B>\Delta_{so})$ as shown in Fig. \ref{fig_1}c has Zeeman spin configuration where two Fermi surfaces are spin polarized along opposite directions, and hence electrons from different Fermi surfaces can have opposite spins, leading to inter-pocket pairing (I) in Fig. \ref{fig_1}f.
Here our argument is based on Fermi surface spin configuration controlled by magnetic field. In the superconducting phase, the Cooper pair momentum can also alter Fermi surface spin configuration. Quantitative calculations taking into account both effects show that the exact phase boundary between I and II at $ T=0 $ is at $\Delta_{so}/B\approx 0.7$.}
The first-order phase boundary between I and II ends at a tricritical point $(\Delta_{so}^{*},B^{*})$ with $\Delta_{so}^{*}\approx 0.7\Delta_0,B^{*}\approx \Delta_0$, where three phases (I, II, N) meet.
Remarkably, near $(\Delta_{so}^{*},B^{*})$ the phase boundary between superconducting phase I and II is a straight line $\Delta_{so}/B\approx 0.7$, closely follows the one between topological and trivial metal in the absence of superconductivity.
This finding demonstrates the direct impact of topology on finite-momentum pairing in spin-orbit-coupled metals.

We further consider low-energy Bogouliubov quasiparticles of the superconducting phase under magnetic field.
As the field strength increases, the energy gap closes and zero-energy quasiparticles form a Fermi surface that is marked different from the normal state Fermi surface.
At $B<\Delta_{so}$ (Fig. \ref{fig_1}d), two Bogouliubov-Fermi segments are formed by inner pocket, while outer pocket is fully gapped.
At $B>\Delta_{so}$ (Fig. \ref{fig_1}f), four Bogouliubov-Fermi segments are formed by both inner and outer Fermi surfaces.
Such Bogouliubov-Fermi segments as shown in Fig. \ref{fig_1}d-f can be measured experimentally via STM spectroscopy \cite{Yuan} or quasiparticle interference.
{In terms of topology, we find Berry phases along zero-energy contours of Bogouliubov quasiparticles are zero, and Bogouliubov-Fermi surfaces are trivial in our model. As a result, within the superconducting phase, topological transitions cannot be realized, and there are either first-order phase transitions ($\Delta_{so}<\Delta_{so}^{*}$) or no phase transitions ($\Delta_{so}>\Delta_{so}^{*}$), as shown in the phase diagram Fig. \ref{fig_2}a.}

%{\bf comment on previous work; general statement about other types of spin-orbit-coupling}

\begin{figure}
\begin{center}
\leavevmode\includegraphics[width=1.\hsize]{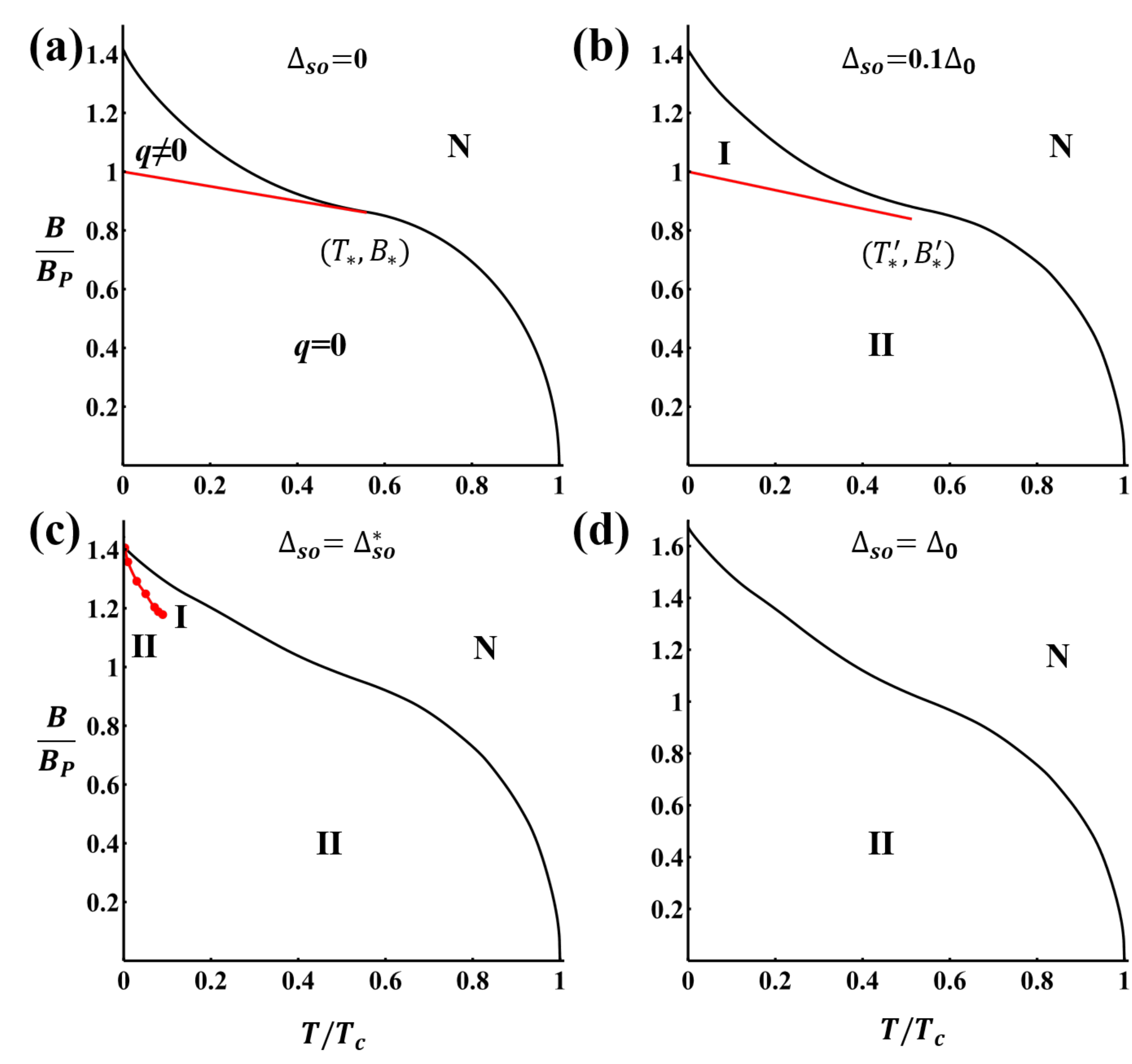}
\end{center}
\caption{Phase diagram in the $T$-$B$ plane with different SOC strengths. Black (red) lines denote second (first) order phase transitions. We set (b) $\mu =10 T_c$ and (c,d) $\mu =50 T_c$ to include DOS asymmetry, which drives the tricritical point $ (T_*,B_*) $ in (a) to a critical point $ (T'_*,B'_*) $ in (b).}
\label{fig_3}
\end{figure}

Our results on finite-momentum superconductivity at $T=0$ has important implications for the global phase diagram as a function of temperature and magnetic field, which is plotted in Fig. \ref{fig_3} for different SOC strengths. Without SOC, phases I (FFLO) and II (BCS) are separated by a first-order phase transition line in the $T$-$B$ plane, which starts at $(T=0, B=B_P)$ and ends at a finite-temperature tricritical point $(T_*=0.56T_c, B_*=0.6\Delta_0)$ where phases I, II and N meet \cite{Maki1,Maki2}, see Fig. \ref{fig_3}a. %We may call $(T_*,B_*)$ as the thermal tricritical point, different from the topological one $(\Delta^*_{so},B^*)$. Notice that thermal tricritical field $B_{*}=0.6\Delta_0$ and topological tricritical field $B^{*}=\Delta_0$ are different.

%Since SOC is non-pair-breaking, the superconducting state at $B=0$ in spin-orbit-coupled metals necessarily has the $\bm q=0$ intra-pocket pairing.
In the presence of SOC, a small field $B\ll\Delta_{so}$ displaces the centers of inner and outer Fermi pockets to opposite momenta $\pm {\bm k}_0 \propto \pm \bm B$. Then, pairing within the inner (outer) Fermi pocket would lead to Cooper pair momentum $\pm {\bm q} \equiv \pm 2 {\bm k}_0$ respectively.
Importantly, due to the difference in DOS on the two pockets, the pairing susceptibilities $\chi(\bm q)$ and $\chi(-\bm q)$ are generally unequal. The larger of the two sets the Cooper pair momentum near the superconducting transition temperature. This argument shows that the Cooper pair momentum $\bm q$ is linearly proportional to $\bm B$ in the weak field regime. The resulting finite-momentum superconductor is characterized by intra-pocket pairing and evolves smoothly out of the BCS state at $B=0$, {and hence is different from the conventional FFLO phase at $B>B_P$}.

%The last term describes the coupling of the Cooper pair momentum and the Zeeman field due to SOC.
The linear coupling between Cooper pair momentum and in-plane magnetic field can be also deduced at a formal level from the Ginzburg-Landau free energy in terms of the real-space order parameter $\psi(\bm r)$, $F=\int d {\bm r} f$ with
\begin{eqnarray}
f=\psi^{*}\alpha \psi + \beta |\psi|^4.
\end{eqnarray}
The coefficient of the quadratic term $\alpha$ can be expanded in powers of the wavevector $\bm q$. Up to fourth order, it takes the following form dictated by symmetry,  %in terms of momentum $\bm q=-i\nabla$
\begin{equation}\label{eq_f1}
\alpha=a_{0}+ a_1 q^2 +a_2 q^4
-(b_0 +b_1 q^2)\bm q\cdot {\bm \Lambda}_{\bm B},
\end{equation}
where ${\bm \Lambda}_{\bm B}$ is an odd-in-$\bm B$ vector.
Here, in addition to even-order terms $a_0, a_1, a_2$,  odd terms $b_0, b_1$ may be  allowed in spin-orbit-coupled systems, which are invariant under {\it joint} rotation  of  Cooper pair momentum and the Zeeman field.  At weak field, ${\bm \Lambda}_{\bm B}\propto {\bm B}$. By minimizing $\alpha$, we find the induced Cooper pair momentum $\bm q$ near normal-superconducting transition is proportional to $\bm B$:
\begin{eqnarray}
\bm q=\frac{b_0}{2a_1} \bm\Lambda_{\bm B} \propto \bm B
\end{eqnarray}

\begin{figure}
\begin{center}
\leavevmode\includegraphics[width=1\hsize]{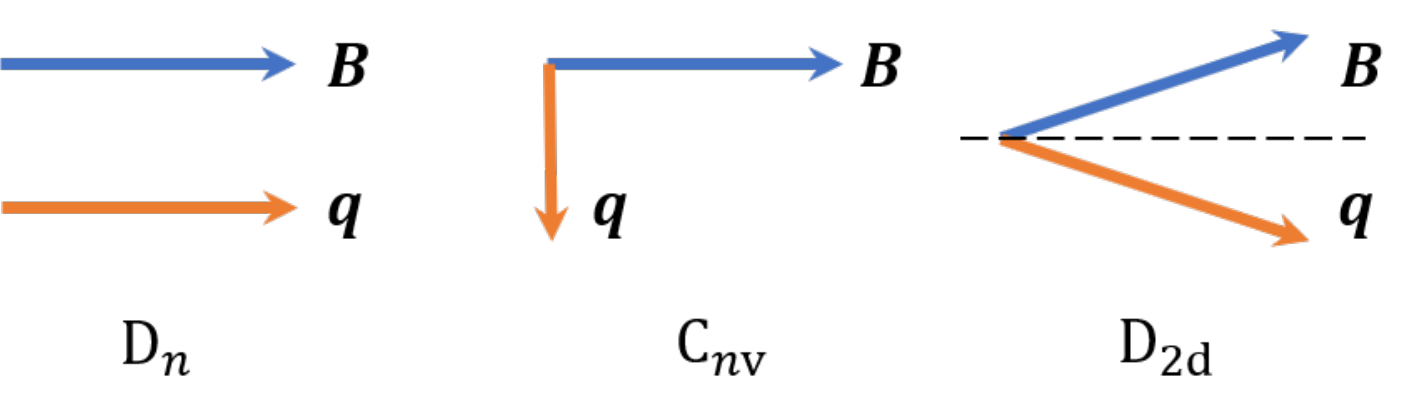}
\end{center}
\caption{Cooper pair momentum $\bm q$ is determined by magnetic field $\bm B$ under different point groups. In D$_{2\rm d}$, directions of $\bm q,\bm B$ form a mirror pair with respect to the mirror plane denoted by a dashed line.}
\label{fig_4}
\end{figure}

Since the new terms are odd under inversion $I:\bm q\to -\bm q,\bm B\to\bm B$ or reflection $ M_z:\bm q\to \bm q,\bm B\to-\bm B $, they exist in systems with broken $I$ and $M_z$, or equivalently in the following 15 point groups:
${\rm D}_{n},{\rm C}_{n\rm v},{\rm C}_{n},{\rm D}_{2\rm d},{\rm S}_{4},{\rm C}_1$ $(n=2,3,4,6)$. The direction of ${\bm \Lambda}_{\bm B}$ and hence the induced Cooper pair momentum depends on the point group symmetry.  %For ${\rm C}_{n\rm v}$ systems with $n=3,4,6, \infty$, $ \bm\Lambda=\bm B\times\hat{\bm z} $.
Fig. \ref{fig_4} shows the direction of $\bm q$ for point groups ${\rm D}_{n},{\rm C}_{n\rm v} $ and $ {\rm D}_{2\rm d} $ respectively. For 2D Rashba systems, the induced Cooper pair momentum at weak field can be obtained by calculating $a_1$ and $b_0$ using BCS theory:
$
{\bm q}=2\alpha_{\rm R}\bm B\times\hat{\bm z}/v_{\rm F}^2
$ \cite{Olga}.
Note that the DOS asymmetry must be included to obtain a nonzero $\bm q$.
%\end{widetext}

By numerically calculating the susceptibility $\chi(\bm q)$ as a function of $B$, we locate the normal-superconducting phase boundary, i.e., the upper critical field curve $B_c(T)$. Without SOC ($b_0=b_1=0$),  the $B_c(T)$ curve is divided into two parts by a tricritical point $(T_*, B_*)$, which corresponds to $a_0 = a_1 =0$ and $a_2>0$. The Cooper pair momentum at the onset of superconductivity changes from $q=0$ at $B<B_*$ to $q= \sqrt{|a_1|/2a_2}$ at $B>B_*$.

When SOC is present, due to $b_0, b_1 \neq 0$ the Cooper pair momentum is already nonzero at weak field. Our microscopic calculation shows that for both small and large Rashba couplings, the Cooper pair momentum at $T_c$ changes smoothly with the field. In other words, there is no tricritical point on the $B_c(T)$ curve. On the other hand, a perturbatively small SOC strength cannot eliminate the strong first-order transition between BCS and FFLO states at low temperature. Therefore, for small SOC strength, we expect the phase diagram shown in Fig. \ref{fig_3}b. Since at $B\neq 0$ the superconducting phases I and II both have finite-momentum Cooper pairs and share the same symmetry, the first-order transition between them starts at $(T=0, B\gtrsim B_P)$ (see Fig. \ref{fig_2}a) and ends at a critical point $ (T'_{*},B'_{*}) $, which is located inside the superconducting phase and away from the $B_c(T)$ curve (Fig. \ref{fig_3}b).

As SOC strength increases, the critical point $ (T'_{*},B'_{*})$ moves to higher field and lower temperature.
At certain SOC strength $\Delta_{so}=\Delta_{so}^{*}$, a zero-temperature tricritical point $(0,B^{*})$ arises (Fig. \ref{fig_3}c), a direct result of the topological transition of normal state Fermi surface.
Near the zero-temperature tricritical point $(0,B^{*})$, Eq. (\ref{eq_f1}) also applies.
In a small range of SOC strength $ \Delta^{*}_{so}\leqslant\Delta_{so}\leqslant \Delta^{**}_{so}\approx 1.1\Delta^{*}_{so} $, only phase II exists at zero temperature, while a short first-order line between phase I and II remains at finite temperature. Finally, for  $\Delta_{so}>\Delta^{**}_{so}$, the entire superconducting region is phase II with inter-pocket pairing \cite{Olga,Samokhin1,Samokhin2,Patrick,Gorkov,Agterberg,Kaur,Aoyama,Fulde} as shown in Fig. \ref{fig_3}d.
Putting all these results together, we arrive at a global phase diagram of 2D superconductors under an in-plane magnetic field, for different SOC strengths.

\textit{Conclusion}---
To conclude, we show that the interplay between SOC and Zeeman effect leads to new normal and superconducting phases in metals. The Fermi surface transition in normal phase is of topological origin and drives a first-order phase transition within the finite-momentum superconducting state. Depending on the SOC strength, we find different phase diagrams of 2D superconductors under in-plane magnetic field.

\textit{Acknowledgment}---
We thank Joe Checkelsky, Aravind Devarakonda and Susan Stemmer for stimulating discussions.
This work is supported by DOE Office of Basic Energy Sciences, Division of Materials Sciences and Engineering under Award DE-SC0010526. LF is partly supported by the Simons Investigator award from the Simons Foundation.

\end{document}